\documentclass{article}
\usepackage{fortschritte}
\def\beq{\begin{equation}}                     %
\def\eeq{\end{equation}}                       %
\def\bea{\begin{eqnarray}}                     
\def\eea{\end{eqnarray}}                       
\usepackage{amssymb}

\newcommand{\eqref}[1]{(\ref{#1})}

\newcommand{\id}{\hbox{1\kern-.27em l}}
\newcommand{\sid}{\hbox{\scriptsize1\kern-.27em l}}

\newcommand{\we}{\kern-.1em\wedge\kern-.1em}
\newcommand{\scal}{\kern-.13em\cdot\kern-.13em}

\newcommand{\II}{I\kern-.09em I}

\newcommand{\Z}{\mathbb{Z}}
\newcommand{\R}{\mathbb{R}}

\newcommand{\spa}{\ \ ,\ \ \ \ }

\begin {document}
\def\titleline{
%
%
%
Black Holes and Black Strings on Cylinders    
}
\def\email_speaker{
{\tt obers@nbi.dk
%
%
speaker@engine.institute.country             
}}
\def\authors{
%
%
%
%
%
T. Harmark \1ad , N.A. Obers \2ad            
}
\def\addresses{
%
%
%
%
\1ad                                        %
 Jefferson Physical Laboratory \\
Harvard University,
 Cambridge, MA 02138, USA

\2ad
 The Niels Bohr Institute \\
Blegdamsvej 17, DK-2100 Copenhagen \O, Denmark
}
\def\abstracttext{
%
%
We review the recently discovered ansatz that describes
non-extremal charged dilatonic branes of string/M-theory
with a transverse circle. The ansatz involves a new coordinate
system that interpolates between the spherical and cylindrical case,
 and reduces the equations of motion to a set of equations
 on one unknown function of two variables.
 The function is independent  of the charge, so that the ansatz can
also be used to construct neutral black holes on
cylinders and near-extremal charged dilatonic branes with a
transverse circle. The construction enables us to
argue that, for sufficiently large mass, there exists a
 neutral solution that breaks translational invariance
in the circle direction, and has larger entropy than that of the neutral
black string of the same mass.

} \large \makefront
\section{Introduction}

The study of black holes on cylinders $\R^{d-1} \times S^1$
is interesting from many perspectives. While in flat space the
static neutral black hole is uniquely
described by the Schwarzschild solution, black holes on
cylinders exhibit a more interesting dynamics and richer
phase structure, as the radius of the circle introduces
a macroscopic scale in the system.
Moreover, $\R^{d-1} \times S^1$ has a non-trivial topology
in the sense that it is non-contractible
with a non-trivial fundamental group. As a consequence,
the event horizon of black objects may not only have topology $S^{d-1}$
but also $S^{d-2} \times S^1$. Finally, black holes on cylinders
show up in various contexts in string theory as branes on a transverse
circle, their near-horizon limits being  relevant
in the AdS/CFT correspondence and Matrix theory.

If we start with a small mass black hole on a cylinder and raise
its mass, so that the horizon radius increases, one will reach at
some point a critical mass $M_c$ for which the event horizon meets
itself across the cylinder. At present the following scenarios
have been proposed for what happens when this point in the phase
diagram is reached: \newline $\bullet $ The most standard point of
view is that for $M > M_c$ the black will uniformly distribute its
mass along the circle, forming a neutral black string with horizon
topology $S^{d-2} \times S^1$, which is classically stable. The
corresponding transition is first order. \newline $\bullet $ The
second scenario which is put forward in \cite{Harmark:2002tr}
claims the existence of a new solution for $M >M_c$, also with
horizon topology $S^{d-2} \times S^1$,  but which is not
translationally invariant along the circle (see also
Ref.~\cite{Horowitz:2002ym}). The phase transition involved is of
higher order (second or third). This non-uniform solution has
higher entropy than the corresponding black string of equal mass.
Only in the infinite mass limit do we recover the black string.
\newline
$\bullet$ There is another mass scale in the problem $M_{u} > M_c$,
such that for $ M_c < M < M_{u}$ we have a phase with a non-uniform
black string and for $M > M_{u}$ a phase with a uniform black string.
This scenario was put forward in \cite{Kol:2002xz}
motivated by \cite{Horowitz:2001cz,Gubser:2001ac}. The scenario has
been considered further using numerical techniques in
\cite{Wiseman:2002zc,Wiseman:2002ti}.

In this talk we review the evidence for the second scenario above,
as presented in Ref.~\cite{Harmark:2002tr}.
The claim is supported by following an analytic approach
toward constructing explicit solutions for black holes on cylinders
$\R^{d-1} \times S^1$ for $d \geq 4$ %
\footnote{Explicit solutions have been constructed for black holes
on $\R^2 \times S^1$ \cite{Myers:1987rx}. 
However, the methods
used there are highly particular to that case and cannot be
generalized to $\R^{d-1} \times S^1$ for $d \geq 4$.}. As part of
the construction we find a new coordinate system in which we are
able to conjecture a general ansatz for non-extremal charged
dilatonic branes with a transverse circle.
 Note here that we sometimes loosely refer to $p$-branes with
$d$ transverse directions as black holes in a $(d+1)$-dimensional
space-time, since the $p$ world-volume directions can be considered
to be compactified.

We show that the equations of motion imply that the ansatz is in fact fully described
by only one function, and present strong arguments in favor of
consistency of the system.
Moreover, we find that this function is independent of the charge
which means that
we can map the ansatz for non-extremal charged dilatonic
branes with a transverse circle to an ansatz for the near-extremal
limit of those and, moreover,
to an ansatz for neutral black holes on cylinders.
Thus, our construction applies to three classes of black holes on
cylinders: the non-extremal charged dilatonic branes of
String/M-theory with a transverse circle,
the near-extremal limit of these branes and finally
neutral black holes on cylinders.

The existence of the new solutions has the consequence
that for $M > M_c$ we have another black object
with the same horizon topology as the black string.
We argue via our construction that
the new solutions have larger entropy than the black strings.
This means that for $M > M_c$ the black string can gain entropy
by spontaneously breaking the translational invariance and
redistributing its mass according to our new solution.
We have thus found a new instability of the black string for
large mass $M > M_c$ which is not classical in nature.
The argument easily generalizes to a similar statement
for the non- and near-extremal case.

To further put our result in broader context, is also instructive
to consider a very massive black string on a cylinder and imagine
what happens when one lowers its mass. In this case the
corresponding scenarios are as follows: \newline $\bullet$
Gregory and Laflamme \cite{Gregory:1993vy} found that a neutral
black string wrapping a cylinder is classically unstable if its
mass becomes sufficiently small $M < M_{\rm GL}$. Since the
entropy of a black hole with the same mass is larger, it was
conjectured that the black string decays to a black hole. This
involves a first order transition from a black object of topology
$S^{d-2} \times S^1$ to one of topology $S^{d-1}$.
\newline
$\bullet$ Horowitz and Maeda \cite{Horowitz:2001cz}
 argued that an event horizon cannot have a
collapsing circle in a classical evolution. This obstruction for
the black string to change the topology of the event horizon, lead
to the conjecture \cite{Horowitz:2001cz} that there exist new
classical solutions with event horizon of topology $S^{d-2} \times
S^1$ that are non-translationally invariant along the circle
whenever the black string is classically unstable. The black
string thus decays to this non-uniform solution.
Following the investigation of \cite{Gubser:2001ac}
and the further work
of Refs.~\cite{Kol:2002xz,Wiseman:2002zc,Wiseman:2002ti}
one is then lead to the reverse process of
the third bullet above.
 \newline
$\bullet$ Finally we have the scenario advocated in
Ref.~\cite{Harmark:2002tr}, in which for $M >M_c$ there exists a
non-uniform black hole solution while for $M < M_c$ the black hole
solution is found after a higher order phase transition. This is
of course the reverse of the second bullet above.

The focus in the present talk will be directed toward black holes
on cylinders with  large masses $M > M_c$, for which we summarize
the arguments for the existence of solutions
 that are non-uniform and classically stable \cite{Harmark:2002tr}.
Another approach to this problem can be found in
Ref.~\cite{Horowitz:2002ym}, which uses
specially constructed initial data to argue that
certain classes of charged non-extremal $p$-branes have
a stable non-translationally invariant solution with horizons of topology
$S^{d-2} \times S^1$.
\footnote{See also Ref.~\cite{DeSmet:2002fv} for
a Petrov classification of five-dimensional metrics, relevant
for the construction of black holes on $R^3 \times S^1$.}

\section{New coordinates and the ansatz}

Finding solutions of black holes on the $d$-dimensional cylinder
$\R^{d-1} \times S^1$ involves solving highly complicated
nonlinear equations. One way to see this is by considering the covering
space $\R^d$ of the cylinder, on which a black
hole in $\R^{d-1} \times S^1$ is really a one-dimensional array of
black holes. Since the interactions between black holes are in
general non-linear, the geometry is very complicated once the
back-reaction is included. Another way to see the complication is
to note that neither the spherical symmetry nor the cylindrical
symmetry applies in general for such a black hole solution.
As a consequence one is forced to consider a solution with
functions that depend on two coordinates rather than one.

Another essential ingredient in finding an appropriate ansatz
is the requirement that the solution should interpolate between the
usual black brane with transverse space $\R^d$,
 which is a good description at
small mass, and the black brane smeared on the transverse circle,
which is a good description at large mass.
We furthermore demand that the solution should reduce to
the extremal charged dilatonic $p$-branes with
transverse space $\R^{d-1} \times S^1$ for zero temperature.
In order to capture these features in an ansatz
we must therefore find an appropriate coordinate system that can be used
in both the small and large mass limits and
also for the extremal solution.

In finding this coordinate system we use the extremal dilatonic
$p$-brane with transverse space $\R^d \times S^1$ as a guide.
This background involves the harmonic function
\begin{equation}
H = 1 + \frac{L^{d-2}}{R_T^{d-2}}
F_{d-2} \Big( \frac{r}{R_T} , \frac{z}{R_T} \Big) \spa
F_{2s} (a,b) \equiv \sum_{n\in \Z} \Big( a^2 + (2\pi n + b)^2 \Big)^{-s}
\end{equation}
where $R_T$ is the size of the transverse circle, $L$ is related
to the charge and $(r,z)$ are cylindrical coordinates. The notation
used for the spherical coordinates is $(\rho,\theta)$.
 We then define
a new coordinate $R(r,z)$ by
\begin{equation}
\label{Rdef}
R^{3-d}
= k_d^{-1} F_{d-2} \Big( \frac{r}{R_T} , \frac{z}{R_T} \Big)
 \end{equation}
which has the property that it interpolates between the radial
variable $r$ in cylindrical coordinates for $R \gg 1$
and the radial variable in spherical coordinates $\rho =\sqrt{r^2 + z^2}$
for $R \ll 1$.
Here $k_d$ is a number that can be found in \cite{Harmark:2002tr}.
The coordinate $v$ that similarly interpolates between $z$ for $R \gg 1$
and an angle $\theta$ on the $S^{d-1}$ sphere for $R \ll 1$
is found by imposing that the flat space metric in the
new coordinates $(R,v)$ is diagonal. This leads to the defining
set of  equations
\begin{equation}
\label{pavr}
\frac{\partial v}{\partial r}
= \frac{1}{(d-3)k_d}  \left( \frac{r}{R_T} \right)^{d-2}
\frac{\partial F_{d-2}}{\partial z} \spa
\frac{\partial v}{\partial z}
= - \frac{1}{(d-3)k_d} \left( \frac{r}{R_T} \right)^{d-2}
\frac{\partial F_{d-2}}{\partial r}
\end{equation}
which is integrable as a consequence of the harmonic equation
$\nabla^2 F_{d-2}=0$. Ref.~\cite{Harmark:2002tr} gives an explicit
solution of these equations in terms of a series expansion.
It is straightforward to write down the flat metric on
the $d$-dimensional cylinder, and that of the extremal dilatonic
$p$-brane in terms of the $(R,v)$ coordinates.

In order to formulate our ansatz in the new coordinate system,
we first specify the boundary conditions that we want to impose.
To this end we introduce a general Schwarz- \linebreak schild radius $R_0$ as
the maximal value of the $R$ coordinate on the horizon. The
boundary conditions are then: \newline
{\bf (i)} The solution reduces
to an ordinary black $p$-brane with transverse space
$\R^{d}$ when $R_0 \leq R \ll 1$. \newline
{\bf (ii)} The solution reduces to a black $p$-brane
smeared on the transverse circle when $R \geq R_0 \gg 1$. \newline
{\bf (iii)} For $R_0/R \rightarrow 0$ the solution approaches the
 extremal dilatonic $p$-brane with transverse space $\R^{d-1} \times S^1$.
 \newline
{\bf (iv)} The horizon is located at constant $R$.
The rationale behind this condition is that the equipotential surfaces
of the charge potential are defined by $R$ being constant
and we expect the horizon to be at an equipotential surface%
\footnote{This holds also for spinning brane solutions
(See for example \cite{Harmark:1999xt}).}.
Obviously, the horizon is then defined by the equation $R=R_0$.

Our ansatz for charged dilatonic black holes on $\R^d \times S^1$
is then
 \begin{equation}
\label{metans2}
ds_{D}^2 = H^{-\frac{d-2}{D-2}}
\left[ - f dt^2 + \sum_{i=1}^p (dx^i)^2
 + H R_T^2
\left( f^{-1} A dR^2 + \frac{A}{K^{d-2}} dv^2
+ K R^2 d\Omega_{d-2}^2 \right) \right]
\end{equation}
\begin{equation}
e^{2\phi} = H^{a}
\spa
A_{01 \cdots p}
= \coth \alpha \Big( 1 - H^{-1} \Big)
\end{equation}
\begin{equation}
\label{fandH}
f = 1 - \frac{R_0^{d-3}}{R^{d-3}}
\spa
H = 1 + \frac{R_0^{d-3} \sinh^2 \alpha}{R^{d-3}}
\spa h_d = R_0^{d-3} \cosh \alpha \sinh \alpha
\end{equation}
with two undetermined functions $A(R,v)$ and $K(R,v)$ at this
point. In this form, the background already satisfies the equation
of motion (EOM) of the dilaton and the gauge potential, so that
the only remaining non-trivial EOMs are those for the metric. It
turns out that these are independent of the constant $h_d$ which
is proportional to the charge. We can therefore map our ansatz for
non-extremal charged dilatonic branes with a transverse circle to
an ansatz for neutral black holes on cylinders ($H =1$ in
\eqref{metans2}) as well as near-extremal charged dilatonic branes
with a transverse circle. The boundary conditions will be modified
accordingly. Consequently, the problem of finding solutions of
black dilatonic $p$-brane with transverse space $\R^{d-1} \times
S^1$ can be mapped to that of finding neutral black holes on
$\R^{d-1} \times S^1$, which we focus on in the remainder.

The EOMs are given by $R_{\mu \nu} = 0$, yielding
four non-trivial equations, $R_{RR} =R_{vv} =R_{Rv} =R_{\phi_1 \phi_1} =0$
(see Eq.~(6.1-4) in \cite{Harmark:2002tr} for the explicit form). The
$R_{\phi_1 \phi_1} =0$ equation can be trivially solved for
$A(R,v)$ so that we end up with a system of three differential
equations on $K(R,v)$ containing up to four derivatives.
At this point one may wonder about the consistency of the obtained
system of equations. A detailed analysis in Ref.~\cite{Harmark:2002tr}
shows that through second order in a large $R$ expansion these
three equations are consistent. We regard this as non-trivial
confirmation of our expectation that our ansatz is consistent
\footnote{Moreover, recently it was
shown \cite{Wiseman:2002ti} that the numerical non-uniform black string solution of
\cite{Gubser:2001ac,Wiseman:2002zc} can be put into the ansatz of \cite{Harmark:2002tr}.}.
Moreover, by studying the Newtonian
limit, it was also shown \cite{Harmark:2002tr} that
small black holes on cylinders can be correctly incorporated in the
ansatz as well.

We review here some of the salient features of the perturbative
analysis. First, the boundary condition (iii) above becomes
\begin{equation}
\label{kcond}
K(R,v) \rightarrow K_{(0)}(R,v) \ \ \mbox{for} \ \
\frac{R_0}{R} \rightarrow 0
\end{equation}
where $K_{(0)}(R,v)=\left(\frac{r}{R_T} \right)^2 \left(\frac{F_{d-2}}{k_d}
\right)^{\frac{2}{d-3}}$ is the corresponding function
that enters the flat metric on the $d$-dimensional cylinder.
This means that $K(R,v)$ should have an expansion in powers of
$R_0^{d-3}/R^{d-3}$ with $K_{(0)}(R,v)$ being the zeroth order
term. In particular, for $R \gg 1$ we then impose the boundary
condition
\begin{equation}
\label{leadK}
K(R,v) = 1 - \chi (R_0) \frac{R_0^{d-3}}{R^{d-3}} + \cdots \spa
\chi(R_0) = \left\{ \begin{array}{ccc}
\frac{1}{(d-2)(d-3)} & \mbox{for} & R_0 \ll R_c \\
0 & \mbox{for} & R_0 \gg R_c \end{array} \right. \ .
\end{equation}
Here the $R_0 \ll R_c$ value of $\chi$ was obtained from the
Newtonian limit of small black holes on the cylinder, while the
$R_0 \gg R_c$ value follows by comparison to the black string
solution. By combining periodicity properties in $v$ with the
``expansion parameter'' $e^{-R}$ we have then been able to show
through second order that, given a value of $\chi(R_0)$, $K(R,v)$
is completely determined. In some sense the function $\chi(R_0)$
contains all physical information about the solution, for example
the entire thermodynamics can be derived from it. At present we do
not know the exact solution of the EOMs nor the exact form of
$\chi(R_0$), but we can argue the existence of solutions due to
the perturbative analysis as well as on more general physical
grounds. For $R_0 >R_c$ these must necessarily be
non-translationally invariant along the circle.

\section{Thermodynamics and stability}

Though an exact solution is so far lacking, one can still obtain a
surprising amount of information on our new non-uniform solutions.
First, using the Killing vector $\partial/\partial t$,
we find the surface gravity on the Killing horizon, which turns
out to depend on $ \gamma(R_0) \equiv (A|_{R=R_0})^{-1/2}$.
Evaluation of the EOMs on the horizon yields the satisfying
result that $A(R,v)$ is independent of $v$ on the horizon,
so that we have a well-defined constant temperature of the black
hole. The thermodynamics for the neutral case is then summarized
by
\begin{equation}
\label{neutTS}
T = \gamma(R_0) \frac{d-3}{4\pi R_0 R_T}
\spa
S = \frac{1}{\gamma(R_0)} \frac{\Omega_{d-2} 2\pi R_T}{4G_{d+1}}
(R_0 R_T)^{d-2}
\end{equation}
\begin{equation}
\label{neutmass}
M = \frac{\Omega_{d-2} 2\pi R_T}{16 \pi G_{d+1}}
(d-3) (R_0 R_T)^{d-3} \left[ \frac{d-2}{d-3} - \chi(R_0) \right] \ .
\end{equation}
Note that the first law of thermodynamics then implies a relation
 between $\gamma(R_0)$ and $\chi(R_0)$ which was utilized in
 Ref.~\cite{Harmark:2002tr} to examine the consequences of
 a smooth interpolation between small and large black hole
 solutions on cylinders. In fact, we expect this relation
 to specify the physical subspace of solutions by determining
 the $R_0$ dependence of $\chi$.

Further knowledge of $\chi (R_0)$ can be obtained using
again the Newtonian limit. One considers  static matter with both
a mass density and negative pressure, which corresponds to the
binding energy of the black hole on the cylinder. The analysis
shows that $\chi$ can be interpreted directly in terms of
the binding energy relative to the mass of the object. Then
using the physical fact that the binding energy increases with $R_0$ we have
shown that $\chi(R_0)$ is a monotonically
decreasing function, interpolating between the extremal values
given in \eqref{leadK}, and hence always positive.
By comparing the thermodynamics \eqref{neutTS},
\eqref{neutmass} to that of the
neutral black string it then follows that $\chi (R_0 ) \geq 0$
implies the entropy inequality
\begin{equation}
S(M) > S_{\rm str} (M)
\end{equation}
showing that the entropy of the non-translationally invariant
solution is larger than that of the black string. The reason for
comparing the entropies at a given mass, i.e. working in the
microcanonical ensemble, is that we want to check whether it is
thermodynamically favorable for a black string of a given mass to
redistribute itself into the non-translationally invariant
solution in order to gain entropy. It thus follows that this is
{\sl always} favorable, signaling a new instability of the black
string which is quantum mechanical in nature.

\section{Further comments}

There are many issues open for further study, among which we mention
a few. First it would
naturally be very interesting if it were possible to obtain
exact forms of our new solutions by explicitly solving the
equations. Numerical study would similarly be important in order
to further check some of our assumptions. More generally,
it would be desirable
to have a better understanding
of the space of solutions for black holes on cylinders, as well
as the change of topology of the horizon.  It would furthermore be
illuminating to study more precisely the nature of the
quantum instability that causes the black string to decay to
our new solutions.
There are also many
applications in string theory of these new non-uniform solutions.
In particular the application to thermal little string theory
(LST) was started in Ref.~\cite{Harmark:2002tr}, one of the aims
being further insight into the thermodynamic properties
of LST near the Hagedorn temperature
\cite{Maldacena:1997cg,Harmark:2000hw}.
Finally, the work presented here could also be relevant
in the considerations of
\cite{Gubser:2000mm} about stability of near-extremal branes.

\vskip .5cm
{\bf Acknowledgement} NO would like to thank the organizers of the
35th Ahrenshoop meeting for an interesting workshop and the
invitation to present this work.

\providecommand{\href}[2]{#2}\begingroup\raggedright\endgroup

\end{document}